# 4MOST Consortium Survey 8: Cosmology Redshift Survey (CRS)

Johan Richard[1]
Jean-Paul Kneib[2]
Chris Blake[3]
Anand Raichoor[2]
Johan Comparat[4]
Tom Shanks[5]
Jenny Sorce[1,6]
Martin Sahlén[7]
Cullan Howlett[8]
Elmo Tempel[9,6]
Richard McMahon[10]
Maciej Bilicki[11]
Boudewijn Roukema[12,1]
Jon Loveday[13]
Dan Pryer[13]
Thomas Buchert[1]
Cheng Zhao[2]
and the CRS team

[1] Centre de Recherche Astrophysique de Lyon, France
[2] Laboratoire d'astrophysique, École Polytechnique Fédérale de Lausanne, Switzerland
[3] Centre for Astrophysics and Supercomputing, Swinburne University of Technology, Hawthorn, Australia
[4] Max-Planck-Institut für extraterrestrische Physik, Garching, Germany
[5] Department of Physics, Durham University, UK
[6] Leibniz-Institut für Astrophysik Potsdam (AIP), Germany
[7] Department of Physics and Astronomy, Uppsala universitet, Sweden
[8] International Centre for Radio Astronomy Research/University of Western Australia, Perth, Australia
[9] Tartu Observatory, University of Tartu, Estonia
[10] Institute of Astronomy, University of Cambridge, UK
[11] Sterrewacht Leiden, Universiteit Leiden, the Netherlands
[12] Toruń Centre for Astronomy (TCfA), Nicolaus Copernicus University, Poland
[13] University of Sussex, Brighton, UK

The 4MOST Cosmology Redshift Survey (CRS) will perform stringent cosmological tests via spectroscopic clustering measurements that will complement the best lensing, cosmic microwave background and other surveys in the southern hemisphere. The combination of carefully selected samples of bright galaxies, luminous red galaxies, emission-line galaxies and quasars, totalling about 8 million objects over the redshift range $z$ = 0.15 to 3.5, will allow definitive tests of gravitational physics. Many key science questions will be addressed by combining CRS spectra of these targets with data from current or future facilities such as the Large Synoptic Survey Telescope, the Square Kilometre Array and the Euclid mission.

## Scientific context

A wide variety of cosmological observations suggest that, in the standard interpretation, the Universe has entered a phase of accelerating expansion propelled by some form of dark energy. The physical nature of dark energy is not yet understood, and may reflect the general-relativistic nature of structure formation, new contributions in the matter-energy sector, or new fundamental theory, such as modifications to gravitational physics on cosmic scales. Past studies of the effects of dark energy have particularly focused on mapping the expansion history of the Universe, for example, using baryon acoustic oscillations (BAO) as a standard ruler or Type Ia Supernovae as standard candles. These probes have yielded important constraints on the homogeneous expanding Universe, including ~ 1% distance measurements and a ~ 5% determination of the equation of state of dark energy. Future surveys, for example by the Dark Energy Spectroscopic Instrument (DESI)[1] or Euclid[2] will improve these distance constraints to sub-percent measurements in narrow redshift bins.

However, in order to distinguish between the different possible manifestations of dark energy, these measurements of expansion must be supplemented by accurate observations of the gravitational growth of the inhomogeneous clumpy Universe. There are several important signatures of gravitational physics which may be used for this purpose, including the peculiar motions of galaxies or clusters and the patterns of weak lensing imprinted by the deflections of light rays from either distant galaxies or the cosmic microwave background (CMB). These probes rely, to a significant degree, on the cross correlation of imaging datasets with spectroscopic redshift information in order to construct the most precise available observational tests of gravity, and mitigate the most significant systematic effects that limit the efficacy of these tests, such as the calibration of photometric redshifts and galaxy bias. Spectroscopy of the southern hemisphere is vital to enable these advances; there is currently no existing large-scale southern hemisphere redshift survey beyond the local Universe. DESI will survey the northern sky, and the future Taipan Galaxy Survey[3] and Euclid satellite will map structure in the redshift ranges $z < 0.2$ and $1 < z < 2$, respectively, missing out the $0.2 < z < 1$ interval which is key for tracing the physical effects of dark energy. Moreover, current deep imaging from the Dark Energy Survey (DES)[4] and Kilo-Degree Survey (KiDS)[5], future imaging by the Large Synoptic Survey Telescope (LSST)[6], CMB Stage 4 experiments, and future radio surveys by the Square Kilometre Array (SKA) and its precursors, MeerKAT and the Australian Square Kilometre Array Pathfinder (ASKAP), will all concern the southern hemisphere. Southern-hemisphere spectroscopic follow-up with 4MOST is critical for successfully completing the multiple science cases for these facilities.

The 4MOST CRS will make a fundamental contribution to tests of gravitational physics by constructing a unique redshift-space map of the large-scale structure for ~ 8 million galaxies and quasars in the southern hemisphere out to redshift $z$ = 3.5. This map will be cross-correlated with complementary current and future datasets to carry out key cosmological tests. The area of overlap between CRS spectroscopy and lensing-quality deep imaging is about three times that currently planned for DESI, thus enabling compelling and competitive science.

## Specific scientific goals

### Testing gravitational physics with overlapping lensing and spectroscopy
Weak gravitational lensing and galaxy peculiar velocities imprinted in redshift-space distortions are complementary observables for testing the cosmological model because they probe different combinations of the metric potentials.

The overlapping datasets created by the 4MOST CRS are particularly beneficial for these tests (Kirk et al., 2015) because: (1) they allow for the additional measurement of galaxy-galaxy lensing, which is subject to a lower level of systematics than cosmic shear; (2) measurements of quasar magnification bias can be compared with these other lensing measurements and redshift-space distortion analyses in the same volume; (3) imaging can mitigate key redshift-space distortion systematics by constraining galaxy bias models; and (4) the same density fluctuations generate both the lensing and clustering signatures, thus potentially reducing statistical uncertainties.

### Source redshift distributions via cross-correlations

Weak gravitational lensing is one of the most powerful and rapidly developing probes of the cosmological model, being particularly advanced in the southern sky thanks to imaging surveys such as KiDS, DES, and LSST. A principal source of systematic error for cosmic shear tomography is the calibration of the source redshift distribution which enters the cosmological model. Different methods of calibrating this distribution exist and usually they require spectroscopic overlap, which should, however, be deep enough. The planned 4MOST galaxy and quasar redshift surveys will allow for this calibration to be accomplished up to high redshifts for all overlapping imaging surveys in the southern sky.

### Synergies with CMB experiments

As the only planned large southern spectroscopic survey at intermediate redshifts, 4MOST is uniquely positioned for synergies with CMB Stage 4 experiments mapping the CMB across the southern hemisphere with unprecedented resolution and accuracy. The CMB contains a wealth of information about the late-time cosmic evolution through its interactions with the large-scale structure. Of particular importance are the Sunyaev-Zel'dovich and integrated Sachs-Wolfe effects, and weak gravitational lensing of the CMB. CRS will provide growth-rate measurements by allowing cross-correlation with spectroscopically confirmed targets.

## Auxiliary science

### Large-scale structure mapping

CRS will offer a spectroscopic view of both large and small scales of the cosmic web. In particular, its high galaxy number density will allow structural studies of voids down to relatively small scales over a wide redshift range. At larger scales, CRS will further enable cosmological distance and effective expansion rate measurements accurate to 1–5% to be made in bins of $dz = 0.1$ up to $z = 3.5$ using galaxies and quasars, complementing DESI BAO measurements in the northern hemisphere. The CRS Ly$\alpha$ survey will exploit the higher spectral resolution compared to DESI (by a factor of almost 2) to measure structure in the Ly$\alpha$ forest down to sub-Mpc scales, allowing new limits to be placed on warm dark matter models as well as high-redshift BAO measurements (for example, Bautista et al., 2017). Combined with chronometric measures of the effective expansion rate, these BAO distances will also provide tests of average curvature and effective expansion rate consistency (for example, Clarkson et al., 2008), to test the standard hypothesis that comoving space is rigid (Roukema et al., 2015).

### Synergies with other surveys

Cross-correlation of large-scale H I intensity maps across the southern sky with optical spectroscopy will allow the evolution of the neutral hydrogen content of galaxies to be mapped in detail, paving the way to surveys with the SKA (Wolz et al., 2017). CRS cross-correlations with overlapping optical and eROSITA X-ray imaging will allow us respectively to measure the effect of quasar feedback on the local clustering environment and to investigate novel routes to cosmological parameters (for example, Risaliti & Lusso, 2018). CRS, in conjunction with the 4MOST TIDES Survey (Survey 10; Swann et al., p. 58), can map the host-galaxy redshifts of a significant population of SNe discovered by LSST, allowing precise gravitational tests using peculiar velocities (Howlett et al., 2017). CRS will also be a valuable tool to follow up the numerous galaxy-galaxy strong lensing events found by Euclid and LSST (Collett, 2015), which can be used as probes for the dark matter distribution at galactic scales.

## Science requirements

– The minimum survey area needed is 6000 square degrees. The minimum survey area for photometric redshift calibration is 1000 square degrees.
– The minimum required target densities for each target category (Bright Galaxies — BG, Luminous Red Galaxies — LRG, Emission-Line Galaxies — ELG, Quasars — QSO) are defined such that clustering measurements are not limited by Poisson noise.
– We require a spectroscopic success rate (SSR) > 95% for BGs, > 75% for LRGs, > 80% for ELGs and > 50% for QSOs.

The survey area is required to be as wide as possible, a requirement driven by carrying out the best measurements and covering all of the existing high-quality imaging in the southern hemisphere from DES and KiDS. The minimum area of 6000 square degrees (current baseline at 7500 square degrees) is based on ensuring a much wider area (and therefore a strong impact) for CRS compared to the planned overlap area of 3000 square degrees for DESI at $z < 0.7$, where targets have the strongest galaxy-galaxy lensing signal and are best placed to lens sources in DES and KiDS. The latter two requirements are there to ensure sample-variance-limited measurements on large scales and for the efficiency of the survey; these are based on previous experiments with similar target types (for example, eBOSS).

## Target selection and survey area

Cross-correlation with deep lensing and CMB surveys motivates the use of LRG as the most efficient tracers of large-scale structure with the maximal lensing imprint. These targets should span a range of redshifts where the lensing geometry is most efficient and cosmological physics is dark-energy dominated, i.e., $z < 0.7$. Photometric redshift calibration by cross-correlation requires full redshift coverage to the limit of the source sample, where each target class covers at least 1000 square degrees (Newman et al., 2015).

Table 1. Properties of each target category in CRS.

| Name | z | Selected (AB) magnitude range | R-band (magnitude [AB]) | Sky area (deg²) | Density (deg²) | Colour selection | Redshift completeness | Number of targets (10⁶) |
|---|---|---|---|---|---|---|---|---|
| BG | 0.15–0.4 | $16 < J < 18$ | $20.2 \pm 0.4$ | 7500 | 250 | $J-Ks, J-W1$ | 95% | 1.88 |
| LRG | 0.4–0.7 | $18.0 < J < 19.5$ | $21.8 \pm 0.7$ | 7500 | 400 | $J-Ks, J-W1$ | 75% | 3.00 |
| ELG | 0.6–1.1 | $21.0 < g < 23.2$ | $23.9 \pm 0.3$ | 1000 | 1200 | $g-r, r-i$ | 80% | 1.20 |
| QSO | 0.9–2.2 | $g < 22.5$ | $22.2 \pm 0.7$ | 7500 | 190 | $g-i, i-W1, W1-W2$ | 65% | 1.43 |
| QSO-Lyα | 2.2–3.5 | $r < 22.7$ | $22.2 \pm 0.7$ | 7500 | 50 | $g-i, i-W1, W1-W2$ | 90% | 0.38 |

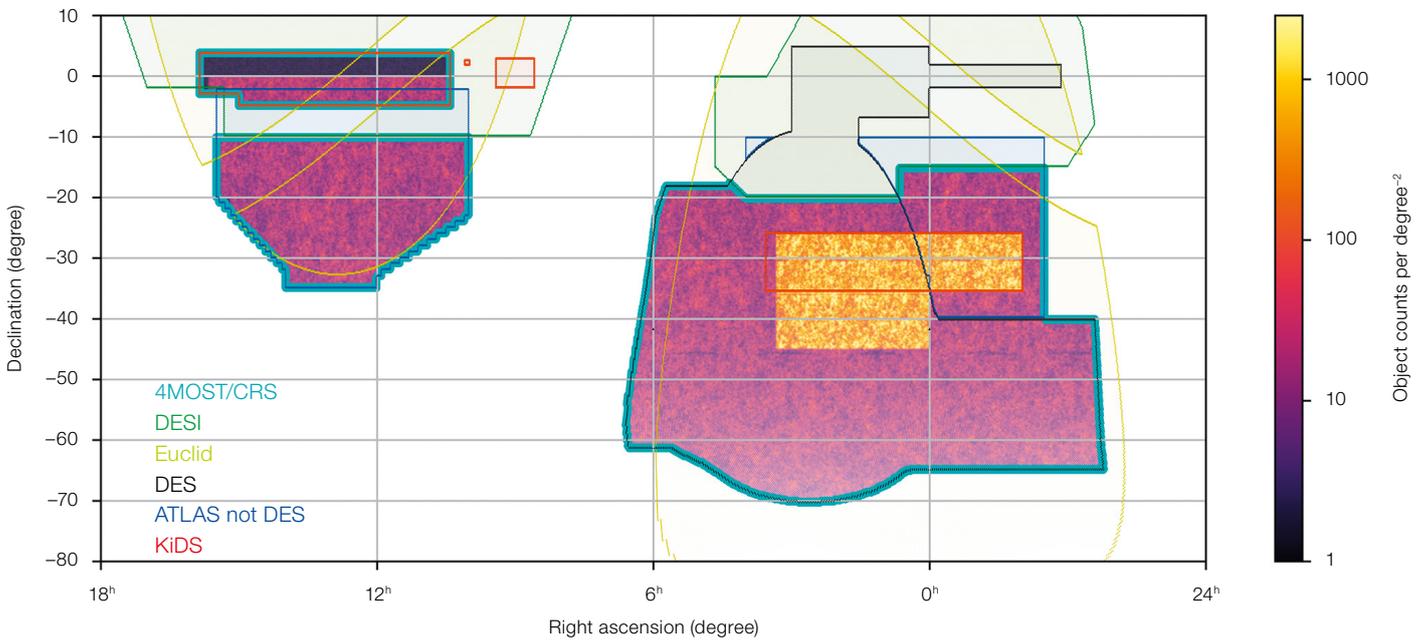

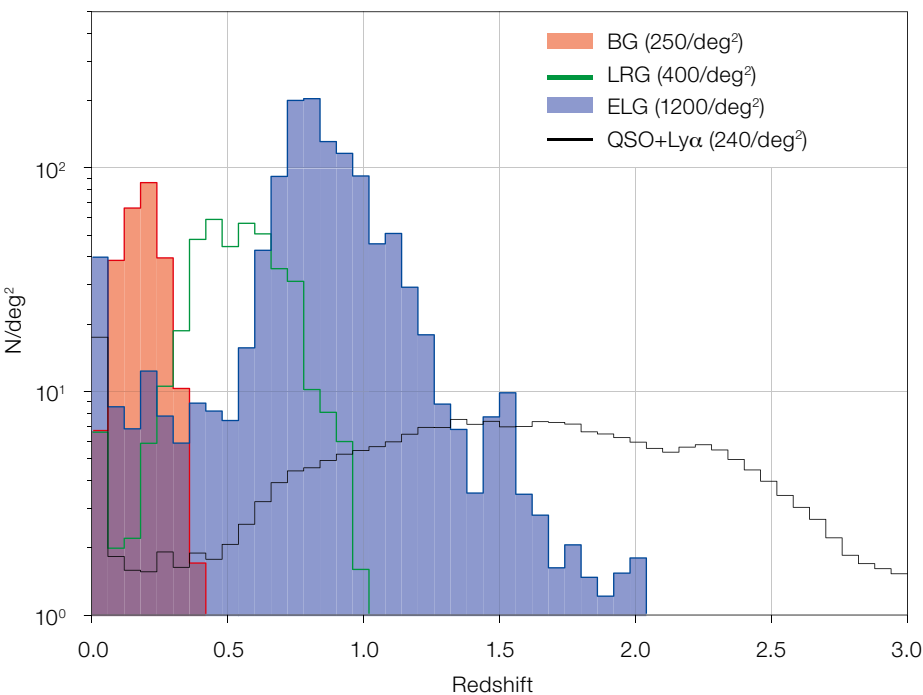

Figure 1 (above). Footprints of the discussed imaging surveys and target densities from mock catalogues. The CRS area (7500 square degrees), demarcated by a thick cyan line, consists of DES and VST-ATLAS excluding DESI and of the two main KiDS regions. The 1000 square degrees covering ELGs is shaded in yellow, with a higher target density.

Figure 2 (left). The expected redshift distributions for the different tracers. These are obtained by applying our target selection on real data, using the HSC photometric redshifts for the BG/LRG/ELG, and using SDSS DR14 spectroscopic redshifts for the QSO.

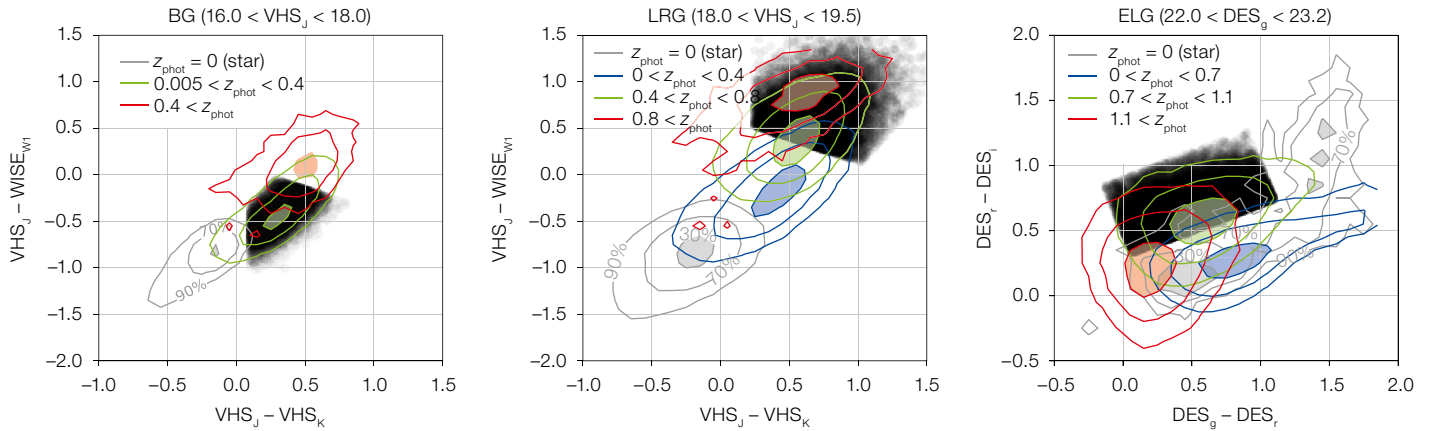

Figure 3. Colour selection for the BG (left), LRG (middle), and ELG (right), using real data (VHS, DES, and the CFHT Legacy Survey photometric redshifts). For each tracer, we display the typical loci of the objects passing the magnitude cut reported in the title: grey contours are for stars, blue/green/red contours are for objects with redshifts lower/within/higher than the aimed redshift range; our selections are shown using black semi-transparent dots.

Targets from the CRS are therefore divided into the following subcategories: BG, LRG, ELG, QSO, including quasars probed through their Lyman-forest at $z > 2.2$ (QSO-Ly$\alpha$). This allows the survey to cover targets at all redshifts from $z = 0$ to $z = 3.5$ (Figure 2). Table 1 summarises the main properties of the magnitude and colour selections.

There are two main survey regions: one larger area of 7500 square degrees for BG, LRG, QSO and QSO-Ly$\alpha$ targets, and a smaller region of 1000 square degrees for ELGs (included in the larger one). The baseline sky area (7500 square degrees) for CRS is constructed by combining the DES, KiDS and VST-ATLAS area which are not covered by DESI (Figure 1). The 1000-square-degrees area for ELG targets is chosen within the best quality imaging region (KiDS-S and DES, Figure 1). There is almost no overlap in ELG targets with the 4MOST WAVES Survey (Driver et al., p. 46), which targets lower redshift sources.

To achieve the 4MOST CRS science goals, it is important to reach a sufficiently large target density in each target category. This density directly translates into a magnitude range in the photometric selection. In addition, colour selections are applied to each target based on empirical regions in the colour-colour diagrams. The colour selections are based on the availability of the relevant filters in the imaging data contained in each region (combining DES as well as the VISTA Hemisphere Survey [VHS] and WISE). The selections foreseen are tuned to obtain the desired target density, maximising the fraction of targets in the desired redshift range and favouring a certain type of objects (red for BG and LRG, blue for ELG, see Figure 3).

## Spectral success criteria and figure of merit

We use the following spectral success criteria to estimate the usefulness of a given target to achieve our science goals:
– **BG and LRG**: median signal-to-noise S/N > 1 per Å in continuum region 4000–8000 Å.
– **ELG**: S/N > 0.5 per Å in continuum region near 6700 Å or 9000 Å.
– **QSO low-z**: S/N > 1 per Å in continuum region near 6700 Å.
– **QSO Lyman-alpha**: S/N > 0.1 per Å in Lyman-alpha forest.

These spectral success criteria are very similar to the ones used for the eBOSS survey (for example, Comparat et al., 2016) and correspond to our goal of reaching a certain redshift completeness at the faintest magnitudes (Table 1).

The figure of merit accounts for the achieved surface density of successful targets and its homogeneity over a large area, which is the main criterion for high accuracy in clustering, as well as the high total number of targets $N$. Each part is equally accounted linearly in the figure of merit calculation.

#### Acknowledgements

We acknowledge support from the French Programme National Cosmologie Galaxies (PNCG), the ERC starting Grant 336736-CALENDS and the ERC advanced Grant 740021-ARTHUS.

#### Links

[1] Dark Energy Spectroscopic Instrument (DESI): www.desi.lbl.gov
[2] Euclid: https://www.euclid-ec.org
[3] Taipan Galaxy Survey: https://www.taipan-survey.org
[4] Dark Energy Survey (DES): https://www.darkenergysurvey.org
[5] Kilo-Degree Survey (KiDS): http://kids.strw.leidenuniv.nl
[6] Large Synoptic Survey Telescope (LSST): https://www.lsst.org